# A possible role of salt-induced intermediates in the liquid-liquid phase transitions of globular protein dispersions

S.P. Rozhkov*, A.S. Goryunov

Institute of Biology, Karelian Research Center RAS,

Pushkinskaya 11, 185910 Petrozavodsk, Russia

**Abstract**

A probable model of the liquid-liquid (L-L) type phase transitions in water-salt dispersions of native (N*, N) globular proteins in the temperature range between thermal (D*) and cold (D) denaturation has been proposed. Protein intermediates (I, I*) arising as a result of the ion non-equilibrium (de)sorption in the process of D↔N and D*↔N* transitions are assumed to be involved in the L-L micro-phase transition forming clusters and fibrils in the main phase of N and N* proteins. Thus, they compensate for their excess chemical potential (ChPot) caused by unbalanced distribution of adsorbed salt ions in protein structure as compared to N-protein. A temperature model is proposed for the behavior of ChPots ($\Delta\mu^i$) of various states of the protein: low-temperature i = D, N, I and high-temperature i = D*, N*, I*, as well as the transition between them. On this basis, the temperature dependence of the solvent ChPot $\Delta\mu_1$ is proposed. The relationship between the extreme values of $\Delta\mu_1$ and the temperatures of the critical points (upper and lower critical solution temperatures) of L-L transitions is established. The reasons for the nonideal behavior of osmotic pressure at different temperatures in water-salt protein dispersions are discussed on the basis of the phase diagrams.

**Highlights**

Some types of phase diagrams on temperature for protein dispersion are proposed

Liquid-liquid phase transitions have been associated with protein intermediates

Nonlinear behavior of proteins chemical potential determines nonideal osmotic pressure

* e-mail: rozhkov@krc.karelia.ru

**Introduction**

The issues of an equilibrium and phase transitions (PT) of the first order between the native (N) and denatured (D) states of the protein have been and remain the subject of intensive research in connection with the problems of protein folding and self-organization [1]. At the same time, liquid-liquid (L-L) PTs in globular protein dispersions, which are usually observed in the range of stabilization of the native form of the protein, also attract a lot of attention. As a rule, the mechanisms of L-L PTs in dispersions of globular proteins are considered out of direct connection with the problems of protein folding [2,3], although internally disordered proteins (IDP) [4] are indispensable participants in L-L PTs in the cytoplasm [5-7].

L-L PTs are associated with the formation of protein clusters (microphases) in the protein concentration (at the limit of solubility) and temperature ranges between the low temperature (D) ↔ (N) and high temperature (D)* ↔ (N)* PT. The role of cluster organization in biosystems is of great importance. Usually it is associated with the phenomena of solute precursors in two-step crystallization [8,9]. Besides, it determines the heterophase organization of membraneless organelles in the cytoplasm [10,11], participates in the regulation of osmotic homeostasis and instant adaptation reactions at the level of macromolecules [12,13], makes for the development of condensation diseases [14,15], food industry technologies [16]. Dense clusters in salt-less solutions contain about a thousand proteins and arise in the temperature range of L-L PTs due to the colloidal interactions: short-range attractive and long-range repulsive potential [2]. Clusters-oligomers 2-10 molecules in size are of a dynamic nature and constantly appear in protein solutions. Their number is proportional to the protein concentration. Clusters of the third type have mesoscopic size of hundred nanometers [2]. The study of the mechanism of the formation of the latter clusters is only at the initial stage [17] and may be associated with the supercritical state of protein dispersion [18].

On the other hand, proteins in saline can be presented as an equilibrium (or non-equilibrium) mixture of macromolecules in different macrostates with different degrees of salt ion sorption. In this case, the binding of ions changes the charge equilibrium of the protein and, as a consequence, changes the interdomain interactions and the conformation of the protein globule.

Solubility of some globular proteins (serum albumin [19], sickle cell hemoglobin [20], α-elastin [21]) is often considered to decrease with increasing temperature, and their solutions have a lower critical solubility temperature (LCST). Another group of proteins (lysozyme, crystallins [22], immunoglobulins [23]) dissolve better with increasing temperature, and their solutions have UCST. In these cases, the amino acid sequence determines the nature of solubility and condensation. Recently, however, there has been progress in the design of internally disordered proteins (IDP), possessing both LCST and UCST, i.e. capable of condensing both during heating and cooling [24]. On the other hand, there are experimental data [25] and theoretical concepts [26-28] indicating that dispersion with both UCST and LCST can be formed by the same protein depending on the microenvironment, i.e. by the composition and pH values of the solution (protein charge), the concentration and type of electrolyte, osmolytes, etc.

The existence of crystalline polymorphism in protein dispersions is associated with the presence of clusters. For example, lysozyme has six crystalline modifications differing in the number of water molecules per amino acid [29]. In this case, the effect of ion binding and hydration on the crystal lattice was found [30]. Proteins can have several tens of binding sites for various anions and cations with different affinities [31,32]. In turn, the ability of ions to bind and their hydration are strongly temperature dependent. This affects their position in the lyotropic series and the selectivity of binding [33-35]. Solubility of polymorphs may sometimes take an opposite temperature trend, from normal to retrograde, or vice versa, as a result of the effects of temperature, pH, and salt concentration [36, 37]. In our model, we thus assume that both UCST and LCST may be characteristic of the same protein dispersion, albeit with a different composition, and we use this as a basis for the proposed phase diagrams.

Internally disordered domains in globular proteins are assumed to promote the formation of intermediates [33]. In this connection, the binding potential of polyvalent metal ions is an important factor in metal-induced conformational changes in protein, leading to the formation of partially folded intermediates. Some metal ions can directly induce fibril formation even at low concentrations. In turn, the presence of specific ion binding sites and the temperature dependence of their affinity are also determined by the amino acid sequence [38]. By varying pH or the concentrations of various salts in solution, it is possible to regulate the charge state of the protein and affect its internal disorder, the number and state of intermediates, and thus the position of L-L PTs on the PD.

Experience in studying of L-L PTs with a critical point (at critical composition and temperature) on relatively simple protein model systems has been accumulated in the literature: water - globular protein [2,3] and water-protein-salt [39]. The dependences of the critical temperatures of L-L PTs on the concentration of proteins, various salts [40] or other osmolytes in dispersion [41] have been experimentally investigated. For the water-protein-salt system, a theoretical condition was obtained for the critical composition corresponding to the critical temperature [39]:

$$\left(\frac{m_2}{m_3}\right) = 2\frac{1+\sqrt{1+\frac{(2+\Delta)^2 \nu^2}{z^2}}}{\nu(2+\Delta)} \quad . \tag{1}$$

Here $m_2/m_3$ is the ratio of the molar concentrations of protein and salt, respectively, $z$ is the charge of the protein, $\nu$ is the amount of electrolyte ions adsorbed by the protein, $\Delta$ is a function of the salt activity coefficient. In this case, $m_2/m_3$ characterizes the critical composition of the system at which a critical PT between the phase represented by the molecular protein solution and the cluster organization of the protein dispersion takes place on the phase diagram (PD). The stable single-phase dispersion forms as a result of this transition. The corresponding value of $m_2/m_3$ implies that a certain number of salt ions should interact with a protein molecule. They must occupy their corresponding sites in the protein structure. In this case, the adjustment can be carried out both by adjusting the temperature to the composition, and the composition to the temperature. In the absence of the required ratio, protein fractions with different numbers of adsorbed ions will be formed differing in conformation, hydration, specific volume, and chemical potential values. Various protein fractions mean the possibility of phase separation as a result of the isolation of molecules with excess potential into a separate phase. We believe that the mechanism of L-L transitions in this case differs from that of L-L transitions in the water-protein system based solely on the DLVO principles.

In this paper we consider protein conformational intermediates (I) and (I)* as fractions of macromolecules with an uneven distribution of adsorbed ions followed by the process of PT (D) ↔ (N) and (D)* ↔ (N)* for these groups of molecules. This distribution shifts for a number of reasons towards higher (or lower) temperatures until thermodynamic equilibrium is reached. Intermediates I present the fraction of the bulk protein that remains in the intermediate conformation between the D and N states, while the rest of the protein acquires N conformation due to D↔N PT. Difference in the effective volumes of N- and I-proteins can lead to a difference in their chemical potentials. In such case, the disturbed equilibrium can be compensated by the release of molecules with a high potential in clusters (a kind of the effect of isothermal distillation). Capillary pressure in such clusters can help equalize chemical potentials in the entire system and establish metastable equilibrium. If the structural-dynamic differences between intermediates and N-molecules decrease, then conditions will arise for a critical PT and the formation of a thermodynamically stable single-phase system of N-molecules.

In this study, we are aimed at justifying a phase diagram, in which the cold N ↔ D and thermal N* ↔D* phase transitions are taken as reference points for the formation of protein fractions (N and I) or (N* and I*) which undergo LLPTs and critical phase transitions along the temperature axis. The phase diagram represents temperature changes in the chemical potentials of various protein states $\mu_i$, where i = D, D*, N, N*, I, I*. In order to relate the state of protein molecules to the state of the protein-solvent system, temperature change in the solvent chemical potential $\Delta\mu_1$ is considered an integral parameter. In turn, this parameter reflects the behavior of the osmotic pressure of the protein dispersion and is able to take into account the contribution of various protein conformational states and associated condensation phenomena to the effective osmotic pressure.

**Modeling**

With increasing temperature at constant pressure, the chemical potential of the substance $\mu$ decreases, since $(\partial\mu / \partial T)_p = -S$, the entropy $S$ being positive [42]. This allows us to qualitatively represent the behavior of the chemical potentials of various states $\mu^i$ of the protein, where $i = D, D^*, N, N^*, I, I^*$ are the corresponding states of the protein. Figure 1 shows such a qualitative diagram in which N and N* protein states are most stable ($\mu^N$ and $\mu^N*$ are lower), and D and D* states are least stable ($\mu^D$ and $\mu^D*$ are higher) in the temperature range between $T_1$ and $T_6$. Accordingly, the concentration of proteins changes towards a more stable state and the content of the D* and D forms becomes negligible. In this case, the potentials of the protein intermediates I and I* take intermediate values, as well as their concentrations. At temperatures $T_1$ and $T_6$ (first-order PT of the macromolecule), the potentials undergo a kink, but their equality ensures the equilibrium of all states of the protein.

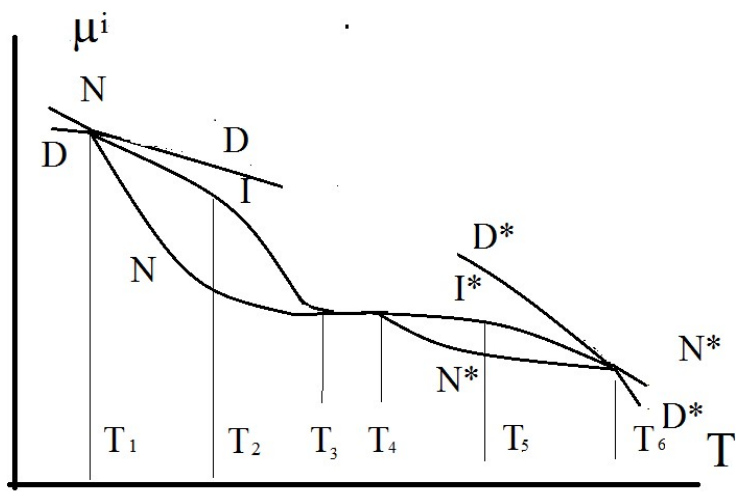

**Fig.1** Temperature dependences of changes in chemical potentials $\mu^i$ of various protein conformers (i = D, D*, N, N*, I, I*). $T_1$ and $T_6$ are N↔D and N*↔D* PTs; $T_2$ and $T_5$ are optimal temperatures for cluster formation; $T_3$ and $T_4$ are critical phase transitions between the phases of native protein and clusters of intermediates. These are UCST and LCST, respectively. There is a supercritical macroscopically single-phase region between $T_3$ and $T_4$.

We assume that the chemical potentials of the various protein states change with increasing temperature, as shown in Fig.1. There are regions near $T_2$ and $T_5$, with the most significant difference between the potentials of the N-I and N*- I* states (Fig. 1). Instability and interactions between macromolecules can be enhanced here. This may lead to a L-L PT, in which molecules with a high potential group into clusters.

Earlier, we theoretically considered a similar case on the assumption that the formation of a cluster or micro-phase is energetically favorable if the polar or charged groups of one protein molecule are located in the region of the nonpolar groups of another protein and affect the state of water in the region of the nonpolar group, increasing its entropy [43]. Moreover, the greater the compensation effect, the fewer protein molecules are needed to form a cluster. It is most likely that proteins with an increased exposure of non-polar groups to water (presumably intermediates) form cluster preferably. Besides, proteins with smaller effective volume than N-proteins, have an increased chemical potential due to Laplace pressure. However, starting from the temperature $T_3$, which is close to the temperature of the critical phase transition, the values of $\mu^N$ and $\mu^I$ equalize and the reasons for the L-L transition disappear, since all molecules become energetically and structurally indistinguishable.

Similar reasoning is valid for the N* and I* conformers in the temperature range below thermal denaturation (T <$T_6$ in Fig. 1). Here the indistinguishability of N*, I* sets in at T <$T_4$. If, in the temperature interval between $T_3$ and $T_4$, the transition between protein conformers occurs along a horizontal line with a common tangent, then this corresponds to a continuous phase transition [42]. In this case, stable conformer N can turn into the stable conformer N*, intermediate I into intermediate I* (and vice versa) by progressive advance of order parameter without nuclei formation.

This suggestion is confirmed experimentally by the data on serum albumin and some of enzymes. It was found [44] that in the temperature range 12–20 ° C, the protein in solution is represented by the only native conformer A, while in the temperature range 22–50 ° C the protein is represented by two native conformers A and B, and the proportion of conformer B increases with temperature. At 58 ° C and above, a spontaneous transformation of A into B occurs.

For solutions of a number of enzymes, it was also shown [45] that there are two native protein conformers with different temperatures of their stabilization. In this case, the high-temperature conformer is supposed to be the state of the molten globule. It has been hypothesized that such a conformation organization is characteristic of all enzymes. According to the data of Foerster energy transfer (smFRET) technique, a large fraction of molecules with a structure close to the high-temperature D-state appears in addition to the N-state, already at room temperatures [46]. A substantial population of the unfolded species is at equilibrium with the folded species at all accessible temperatures [47]. The simultaneous presence of native and denatured molecules leads to a decrease in the thermodynamic stability of the protein solution as a whole [48] and promotes phase separation.

It is not enough to consider only $\mu^i$ for protein molecules in a solvent. The effect of protein molecules on both the $\mu_1$ of the solvent and the $\mu_1$ on the $\mu^i$ of the protein conformers should be also taken into account. Gibbs-Duhem equation is valid for equilibrium system as a whole at any given temperature $\Sigma n_i d\mu_i = 0$. It means that:

$$nd\mu_1 = - (n_N)d\mu^N - (n_I)d\mu^I - (n_D)d\mu^D. \tag{2}$$

The last term can be neglected as $n_D$ is very small. Here n, $n_b$, and $n_h$ are the molar concentrations of the solvent, the bulk solvent, and the solvent involved in hydration: $n = n_b + n_h$. Moreover, $n_h = n^I + n^N$. The values $n^N / n$ and $n^I / n$ characterize the ratio of water moles that are bound to macromolecules in one state and not bound in another. Taking into account that $n^I = n_h - n^N$, $d\mu^i = (\partial \mu^i/\partial T)dT = -S^i dT$, eq. 2 in terms of increments takes the form:

$$\Delta\mu_1 = (n^N/n)(S^N - S^I) \Delta T - (n_h/n)S^I \Delta T \qquad (3)$$

A similar equation is valid for $S^N$ *, $S^I$ *. Thus, the temperature behavior of the $\Delta\mu_1$ will be determined by the difference in the entropies of various states of the protein at the corresponding temperature, i.e. by the slope of the curves $\mu^i$ in Fig. 1 in each temperature range. It is shown in Fig. 1 that $|\Delta\mu/\Delta T|^N > |\Delta\mu/\Delta T|^I$ and $\Delta\mu_1$ decreases in the temperature range ($T_1 - T_2$). On the contrary, $|\Delta\mu/\Delta T|^N < |\Delta\mu/\Delta T|^I$ and $\Delta\mu_1$ increases in the temperature range ($T_2 - T_3$). $|\Delta\mu/\Delta T|^N = |\Delta\mu/\Delta T|^I = 0$ and $\Delta\mu_1$ approaches zero in the interval ($T_3 - T_4$); $|\Delta\mu/\Delta T|^{N*} > |\Delta\mu/\Delta T|^{I*}$ and $\Delta\mu_1$ increases in the interval ($T_4 - T_5$); $|\Delta\mu/\Delta T|^{N*} < |\Delta\mu/\Delta T|^{I*}$ and $\Delta\mu_1$ decreases in the interval ($T_5 - T_6$).

Taking these factors into account eq. 3 allows predicting a substantially non-linear behavior of $\Delta\mu_1$ between $T_1$ and $T_6$. Thus, $\Delta\mu_1$ first decreases and reaches a minimum at $T_2$, and then begins to increase in the range of low temperatures. Accordingly, in the range of high temperatures, $\Delta\mu_1$ first increases with decreasing temperature from $T_6$ to $T_5$, and then begins to decrease. In the temperature range $T_3 - T_4$, these two "half-waves" converge. This is shown in Fig. 2 by phase diagram in the coordinates of $\Delta\mu_1$ of the protein dispersion versus temperature. This curve looks like a fragment of the Tammann elliptic diagram, which can be obtained by integrating the equation $d\Delta G = -\Delta S dT + \Delta V dP - \Delta\Gamma d\mu = 0$ at constant pressure P [49,50]. Here $\Delta\Gamma$ is the factor of preferential hydration [51]. For a more complete correspondence to the perimeter of the ellipse, the points $\Delta S = 0$ and $\Delta\Gamma = 0$ are connected by dotted lines. The last can reflect the presence of protein oligomers and protein fibrils. This diagram makes it possible to compare theoretical estimates of the behavior of $\Delta\mu_1$ (T) with experimental results on osmotic pressure, since the latter is directly coupled with $\Delta\mu_1$.

Alterations in salt concentration can cause changes in $T_1 - T_6$ positions, similar to the effect of salt concentration on the protein stability curve $\Delta G^D_N = G^D - G^N$ and its extreme points, where $(\partial \Delta G^D_N/\partial T)dT = -\underline{\Delta S}dT = 0$.

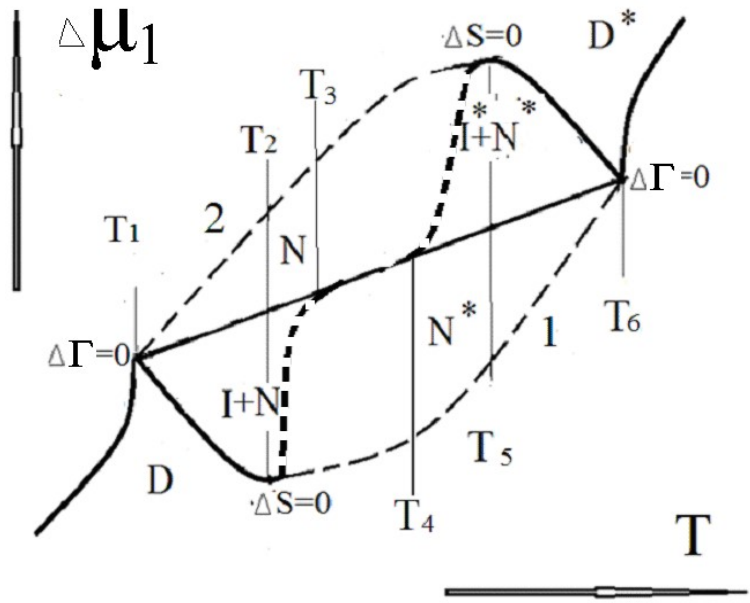

**Fig.2** Phase diagram of globular protein dispersion in the solution ChemPot $\Delta\mu_1$ versus temperature T plane. The extreme points of the elliptical shape curve connect the "neutral" lines $\Delta S = S^I - S^N = 0$ and $\Delta \Gamma = \Gamma^I - \Gamma^N = 0$. D and D * are denatured states, N and N * are native states. I and I * are protein intermediates. I + N, I * + N * are metastable states resulting from L-L transitions. The sign (*) means that the heat and cold forms of the protein may not be identical.

**Discussion**

In the region of cold protein denaturation in Fig.2 $\Delta\mu_1$ increases with temperature up to the point of the first-order PT ($T_1$) associated with the transition to the native state. Moreover, $(\partial\Delta\mu_1/\partial T)\to\infty$ on approaching this point, since $\Delta H = T\Delta S = -T (\partial\Delta\mu_1/\partial T) \neq 0$ is a discontinuous function here. The same is also valid for the high-temperature region. In Fig. 2, temperatures $T_1$ and $T_6$ correspond to N↔D and N*↔D* PTs. If all molecules had the same molar volume and ions binding sites have been equally occupied, then $\Delta\mu_1$ (and osmotic pressure) would follow the line $\Delta\Gamma = 0$. However, it was established that $V^N > V^D$ [52] and $\Delta\Gamma \neq 0$. It is natural to assume that this inequality is also true for protein intermediates I and I*.

Since $\mu^I$ is greater than $\mu^N$ in the range between $T_1$ and $T_3$ (Fig.1), the coexistence of the corresponding fractions can lead to thermodynamic destabilization of the system. However, the dispersion can pass into a metastable state with the formation of clusters of I-molecules, in which the capillary pressure balances the $\mu^I$ excess. The formation of dense protein clusters reflects the predominance of the attraction potential corresponding to negative sign of second virial coefficient of osmotic pressure.

Shown in Fig.2 in bold and dotted lines the two "half-waves" reflect the nonlinear behavior of $\Delta\mu_1$ in accordance with eq (3). At the same time, $\partial(\Delta\mu_1)/\partial T = -(S^I-S^N) = 0$ at the extremal points of $\Delta\mu_1$ ($T_2$ and $T_5$). Keeping in the mind the salt effects, here it can be noted that $\Delta\mu_1$ [51, 18] and the second virial coefficient of osmotic pressure [53] behave in a similar way as in Figure 2, if the salt concentration changes instead of temperature.

Condition for phase separation disappear at temperature $T_3$, since the N-state of protein molecules and the I-state of protein become energetically and structurally indistinguishable. This is the condition for critical transition, which is the extreme case of the first-order PT. It also corresponds to the condition $\Delta S = S^I - S^N = 0$, $\Delta V = V^I - V^N = 0$ and $\Delta\Gamma = \Gamma^I - \Gamma^N = 0$.

PT D*↔N* is also accompanied by the formation of intermediates at high temperatures. However, due to the peculiarities of hydration and interaction with salt ions, conditions are created for the formation of curvilinear fibrils and/or aggregates of linear fibrils as components of a new phase. As shown for human serum albumin [54] and lysozyme [55], a decrease in pH in the presence of salts at room and elevated temperatures can lead to the formation of fibrils. Their kinetic conditions for the formation and their structure substantially depend on the microenvironment. The total interaction potential of such anisotropic elongated formations can become repulsive, and corresponding microphases should be less dense than the main phase. Conditions for phase separation disappear at temperature $T_4$, since $N^*$- state of protein molecules and the state of $I^*$-protein intermediates become energetically and structurally indistinguishable. This is the condition for critical transition, which is the extereme case of the high temperature first-order PT. It also corresponds to the condition $\Delta S = S^N{}^* - S^I{}^* = 0$ ; $\Delta \Gamma = \Gamma^{I*} - \Gamma^{N*} = 0$, and $\Delta V = V^N{}^* - V^I{}^* = 0$.

Thin dotted lines (1) and (2) in Figure 2, connecting the extreme points $\Delta S = 0$ and $\Delta \Gamma = 0$ may correspond, in our opinion, to the contribution to $\Delta \mu_1$ (or the osmotic pressure) of protein oligomers (1) resulting from the destabilization of dense clusters, curvilinear fibrils (2) or, in its turn, ensembles of linear fibrils. Thus, the general form of the PD between $T_1$ and $T_6$, which is close to elliptical, suggests the existence of protein fractions with molecules of different conformations (D, D *, N, N *, I, I *). PD in the form of ellipsoid was obtained earlier in the analysis of D↔N transitions of protein in the model of two states in the pressure P - temperature T coordinates [56]. The PD in the coordinates {T, $\Delta \mu_1$} was shown should have the same shape [49]. This is consistent with the model proposed in Figure 2.

Osmotic pressure is known to behave significantly nonideally depending on the potential of attraction between molecules as well as the interaction of proteins with water and salt ions [57,58]. Accordingly, it should decrease along the corresponding trajectory on the PD in the region of the lower loop in Fig. 2 in the temperature range $T_1$-$T_2$ due to the presence of dense clusters. The sign of the virial coefficient becomes negative on rising salt concentration, and this is associated with the formation of clusters. This is accompanied by low values of osmotic and swelling pressure (when protein gel forms [59,60]). Both nucleation and formation of protein oligomers can take place in protein clusters under conditions of increased molecular density. About 20% of the protein can be arranged in oligomers in the range of neutral pH values [61]. These oligomers are retained partially after the completion of the L-L PT and contribute to the osmotic pressure (dotted line (1) between $T_2$ and $T_6$ in Figure 2). The oligomers will destabilize and the osmotic pressure will rise with rising temperature.

The interaction mechanism changes in the high temperature range. Salt ions are likely to interact differently with protein molecules at low and high temperatures. Anions have been found to adsorb preferentially on the protein at low temperatures (especially for a 1: 1 electrolyte), even in spite of the negative charge of the protein [62]. This seems to be analogous to an increase in the pH of a solution [61]. At high temperatures, the affinity of cations with protein can increase (especially polyvalent ones), which is due to the peculiarities of hydration and entropic effects. This is an analogy with a decrease in the pH of a solution [61]. The surface can be re-charged from (-) to (+) in the process, which can lead to electrostatic repulsion of molecules and a reentrant phase transition with strong adsorption of cations [63]. Experiments on osmotic pressure show that protein denaturation lead to different types of its dynamic behavior and interaction with decreasing and increasing pH [61]. Segmental mobility of the protein increases while lowering pH, which promotes the formation of filaments. Mobility decreases and promotes the formation of protein intermediates and oligomerization while rising pH. Formation of clusters of elongated protein fibrils during the L-L PT at high temperatures from $T_6$ to$T_5$ can cause high positive values of the virial coefficient [64] (the loop between $T_6$ and $T_4$ in Fig. 2). As temperature decreases below the LCST, small curvilinear fibrils remain (dashed line 2 in Figure 2), which are destroyed with a further decrease in temperature. Monomers remain on approaching the cold denaturation region [65] and contribute to the osmotic pressure.

Thus, in the range of native protein states from $T_1$ to $T_6$, the total osmotic pressure and its nonideal values can be determined by the contribution from all possible conformational states of protein molecules, as well as clusters of intermediates, oligomers, various fibrils and aggregates. In Figure 2 this possible total contribution is represented by a set of lines bounded by a second-order figure.

Tammann elliptic diagram [66] is known to describe physicochemical systems with reentrant PT. Water-protein systems are capable of such transitions with changes in temperature (cold D and thermal D * denaturation) D↔N↔D *, concentration of polyvalent electrolyte [67] and hydrostatic pressure [68]. The diagram in Fig. 2 also allows reentrant PTs with changes in the concentration and type of salt, which can influence the temperatures $T_2$ and $T_5$ and the slope of the line $\Delta S = S^N - S^I = 0$. An increase in the salt concentration (ionic strength) leads to a narrowing of the temperature range from $T_2$ to $T_5$ at first, but further growth leads to its expansion. Second, at constant temperature, changing the quality of the solvent transfers the system from the gel region with a low swelling pressure (outside the ellipse), through the protein dispersion (inside the ellipse), to the gel region with a high swelling pressure (outside the ellipse).

Based on Figure 2, an extended PD of the protein dispersion in the temperature-composition coordinates can be proposed (Figure 3). The composition is determined by the ratio of molar protein and salt concentrations $m_2/m_3$ according to equation (1). The diagram takes into account two LL PTs, the possible existence of N and N* conformational states of the native protein and transformations N↔N* and I↔I*, the presence of a large fraction of protein oligomers and ordered aggregates of amyloid-like fibrils. Dashed lines in Fig. 3 show the curves of solubility of protein oligomers and curvilinear fibrils. Both of them can exist in the interval between UCST and LCST as a possible result of destabilization of protein clusters and linear fibrils arising during L-L PTs. Salts significantly affect UCST and LCST. Existing data allows assuming that LCST and UCST first approach each other, then, depending on the ionic strength, begin to diverge with an increase in the concentration of salts. This creates temperature conditions for reentrant PTs. Thus, the implementation of the hypothetical PD in Figure 3 implies varying of the dispersion composition both in the concentration of protein and salts.

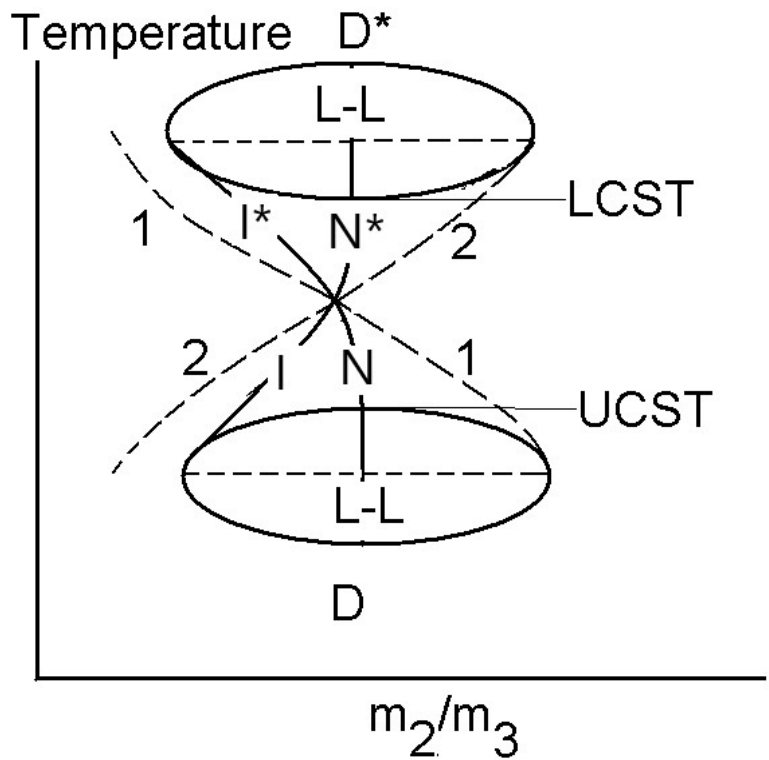

**Fig.3** Phase diagram of globular protein dispersion in the temperature - composition $m_2/m_3$ plane, where $m_2$ and $m_3$ are the molar concentrations of protein and salt, respectively. N and N* are possible conformers of the native protein. Both I and I* conformers can be close to a molten globule in the corresponding temperature range. Dashed curves denote the solubility of protein oligomers (1) and curvilinear fibrils (2) in the supercritical region between LCST and UCST.

In the region of intermediate temperatures, the solubility lines N and N* can intersect. This reflects the increasing conversion of conformer N into N* form as $\mu^{N*} < \mu^{N}$ with rising temperature. On the contrary, the N conformer prevails in the protein solution at low temperatures. Conformers form a true solution in the region to the left of these lines, while the formation of the corresponding condensation states is possible in the region to the right of them. Thus, protein fibrils can appear due to one-stage condensation in the higher temperature range to the right of curve I*. At the same time, protein oligomers and protein I-conformers form condensates at the binodal line of L-L PT. In this case, the formation of both fibrils from oligomers (two-stage fibrillation) and amorphous aggregates-spherulites is possible [69]. On the other hand, their destabilization to the right of the binodal line can lead to the formation of curvilinear fibrils. Such structures have been found for lysozyme [55] . Their solubility decreases with lowering temperature (curve 2 in Fig. 3), while the solubility of oligomers decreases (curve 1 in Fig. 3) with rising temperature.

**Conclusion**

The elliptical phase diagram in pressure-temperature coordinates {P, T} has long been successfully used to clarify the thermodynamic features of cooperative transitions in protein molecules. A diagram of this type also extends to the osmotic pressure versus temperature {Π,T} plane at constant external pressure. Analysis of the behavior of the chemical potentials of the native, denatured states of protein molecules and their intermediates showed that the {Δμ$_1$,T} diagram is capable to reflect the PD of the first-order N↔D, PT of the LL type, as well as critical PT with LCST and UCST occurring at the appropriate temperature and water-salt composition. This diagram allows to take into account the presence in the protein dispersion of oligomers, amorphous and fibrillar aggregates, and gel states, which cause a substantially nonideal behavior of osmotic pressure in the temperature range from cold to heat denaturation. In addition, the PD takes into account the presence in the protein dispersion of reentrant PT, which depends on temperature, concentration of salts and/or other osmolytes.

**Acknowledgements** The study was carried out under state order (project FMEN-2022-0006).
**Compliance with ethical standards**
**Conflict of interest**
The authors declare that they have no conflict of interest.